\documentclass[conference]{IEEEtran}
\IEEEoverridecommandlockouts
\usepackage{cite}
\usepackage{amsmath,amssymb,amsfonts}

\usepackage{multirow}
\usepackage{subfiles}
\usepackage{algorithm}
\usepackage{amsfonts}

\usepackage{algorithmic}
\usepackage{graphicx}
\usepackage{textcomp}
\usepackage{xcolor}
\usepackage{url}
\usepackage{subfigure}

\def\BibTeX{{\rm B\kern-.05em{\sc i\kern-.025em b}\kern-.08em
    T\kern-.1667em\lower.7ex\hbox{E}\kern-.125emX}}
\begin{document}

\title{APVAS: Reducing Memory Size of AS\_PATH Validation by Using Aggregate Signatures
}


\author{Ouyang Junjie, Naoto Yanai, Tatsuya Takemura, Masayuki Okada, Shingo Okamura, Jason Paul Cruz
\thanks{Ouyang Junjie, Naoto Yanai, Tatsuya Takemura and Jason Paul Cruz are with Osaka University. }
\thanks{Masayuki Okada is with JPNIC.}
\thanks{Shingo Okamura is with National Institute of Technology, Nara College.}
}

\maketitle

\begin{abstract}
The \textit{BGPsec} protocol, which is an extension of the border gateway protocol (BGP), uses digital signatures to guarantee the validity of routing information. However, BGPsec's use of digital signatures in routing information causes a lack of memory in BGP routers and therefore creates a gaping security hole in today's Internet. This problem hinders the practical realization and implementation of BGPsec. 
In this paper, we present \textit{APVAS (AS path validation based on aggregate signatures)}, a new validation method that reduces memory consumption of BGPsec when validating paths in routing information. 
To do this, APVAS relies on a novel aggregate signature scheme that compresses individually generated signatures into a single signature in two ways, i.e., in sequential and interactive fashions. 
Furthermore, we implement a prototype of APVAS on \textit{BIRD Internet Routing Daemon} and demonstrate its efficiency on actual BGP connections. Our results show that APVAS can reduce memory consumption by 80\% in comparison with the conventional BGPsec.
\end{abstract}


\section{Introduction} \label{Introduction}

\subsection{Backgrounds}

The \textit{Border Gateway Protocol (BGP)}~\cite{rekhter_border_2006} enables networks, such as an Internet service provider (ISP), to exchange routing information in the level of \textit{autonomous system (AS)} by assigning a unique number to each AS. BGP is also the primary routing protocol used in the backbone of the Internet. 
However, BGP does not verify the validity of routing information being exchanged, and thus an AS always registers routing information received from other ASes as valid even if an adversary manipulates the routing information. 
This fundamental flaw in BGP has caused many incidents that resulted in heavy and serious damages, e.g., Youtube hijacking~\cite{noauthor_youtube_2008} and Ethereum hijacking~\cite{siddiqui_what_nodate}. 
According to some measurement results~\cite{ref:bgp_hijacks}, such a hijack happens about four times a day on average. Therefore, guaranteeing the validity of routing information in BGP is an urgent and significant issue. 


To tackle the aforementioned issue, technologies that guarantee the security of BGP in a \textit{cryptographic} fashion have attracted attention. 
Loosely speaking, these technologies aim to verify the validity of routing information via generation and verification of digital signatures in the routing information. 
Specifically, signatures can be used in two ways, namely, \textit{route origin validation} that only allows advertisements for an IP prefix by the legitimate AS as a prefix owner and \textit{path validation} that guarantees all members of an AS path which is a connection of ASes from a source to a destination. 
Route origin validation is almost consummative by virtue of the practical realizations of RPKI~\cite{RPKI} and ROA~\cite{rfc6482,rfc6483} as related protocols. 
In contrast, path validation has no clear practical realization even though it is instantiated by BGPsec~\cite{lepinski_bgpsec_2017} because its use of digital signatures significantly increases the memory consumption of BGP routers. 
For instance, according to a current estimation~\cite{sriram_rib_2011}, a BGPsec is required to have memory size of several tens of gigabytes. The issue related to the memory size is known as the \textit{memory size problem}.
Moreover, BGPsec lacks experimental evaluations and thus a precise evaluation of the memory size problem remains incomplete. 
 

BGP hijacking has also given rise to hijacking of cryptocurrencies~\cite{AZV2017,EGJ20}, such as Bitcoin, as a new aspect of cybercrime. A recent finding has shown that BGP hijacking~\cite{BGW+19} can only be prevented by the use of BGPsec. Therefore, an essential issue in BGP security can be solved by making BGPsec practical, i.e., \textit{by reducing memory consumption and solving the memory size problem}.

\subsection{Contribution}

In this paper, we present a new path validation protocol named \textit{APVAS (AS path validation based on aggregate signatures)}, which utilizes aggregate signatures~\cite{boneh_aggregate_2003} to combine individual signatures into a single short signature and solve the memory size problem. 
Moreover, we implement a prototype of APVAS on a router daemon software. 
This is a first attempt to measure memory size by the use of state-of-the-art cryptography in actual devices. 
In our experimental environment, APVAS can reduce memory consumption by 80\% compared to the conventional BGPsec. 
We believe that APVAS will become an innovative solution to BGPsec. 


This paper presents two technical contributions. The first contribution is the proposal of a novel aggregate signature scheme named \textit{bimodal aggregate signatures}. 
Aggregate signatures are expected be applicable to BGPsec in cryptographic theory, but the algebraic structures of aggregate signatures in early literature are unsuitable for the current specifications of BGPsec. 
More precisely, \textit{when the original aggregate signatures~\cite{boneh_aggregate_2003,LMRS04} are trivially deployed in BGPsec, either the capability for signature aggregation or security will be lost}. 
In contrast, APVAS can decrease the memory consumption as well as keep the security of BGPsec by the use of bimodal aggregate signatures (See Section~\ref{Bimodal Aggregate Signatures} for details).


The second contribution is the implementation of a prototype of APVAS by extending \textit{BIRD Internet Routing Daemon (BIRD)\footnote{BIRD: \url{https://bird.network.cz/}}}, which is a software that virtualizes a BGP router. 
The lack of experimental evaluation described in the previous subsection is caused by the lack of evaluation tools for BGPsec. 
In contrast, we succeeded in \textit{measuring the performance of APVAS in an actual environment} by leveraging BIRD. 
Although our experiment was conducted on a linear network in a simple fashion, as far as we aware this is the first time that aggregate signatures are evaluated in an actual environment. 
Moreover, by extending our prototype, we can potentially evaluate protocols in future works (See Section~\ref{Experiments} for details). We plan to release the prototype of APVAS to encourage development of BGPsec and future works.



\section{Related Works} \label{Related Work}

The closest works to APVAS are the aggregate path authentication~\cite{bgp:aggr} and APAT~\cite{apat16}. 
These works introduced aggregate signatures~\cite{boneh_aggregate_2003} in BGPsec (and S-BGP~\cite{kent_secure_2000}) to aggregate individually generated signatures into a single short signature. 
However, they did not discuss the serious issue of memory consumption. 
Moreover, they did not provide improvements to aggregate signatures and a prototype implementation on a BGP software, which are our main technical contributions. 
Surprisingly, there is no commercial implementation of the original BGPsec itself~\cite{SM19} although BGPsec fully overcomes security concerns according to recent works~\cite{BGW+19,SBMS20}.



In the past years, BGP security research~\cite{Goldberg14,LSG16,ARTEMIS} aimed to serve a quick response with ``decent'' security by utilizing \textit{filtering} instead of digital signatures. For example, the use of filtering can prevent 85\% of hijacking~\cite{LSG16}, whereas paths can be repaired from a hijacking within a minute~\cite{ARTEMIS}. 
These results have shown how threats are mitigated in the real world. 
However, we mentioned that when the filtering-based approach is used, it is difficult to distinguish whether hijacking is done by a defense against DDoS or by an adversary from the standpoint of a third party. 

\section{Border Gateway Protocol Security Extension (BGPsec)} \label{BGPsec}

In this section, we provide backgrounds on BGP hijacking, path validation, and the main problem of BGPsec. 

\subsection{Motivating Example: BGP Hijacking} \label{BGP Hijacking}

As a motivating example of BGPsec, we explain route hijacking on BGP below. 
As described in Section~\ref{Introduction}, BGP is a protocol used for finding a route to a destination on the entireInternet in per AS unit. 
Each AS is assigned a unique AS number, and BGP uses these AS numbers to distinguish each AS and exchanges routing information via TCP. 
After a TCP connection is established, 
ASes exchange routing information with each other by sending and receiving an update message containing \textit{Network Layer Reachability Information (NLRI)} and the \textit{AS\_PATH}. 

\begin{figure}[tbp]
 \begin{minipage}{0.49\hsize}
 \begin{center}
 \includegraphics[width=.99\textwidth,height=.20\textheight]{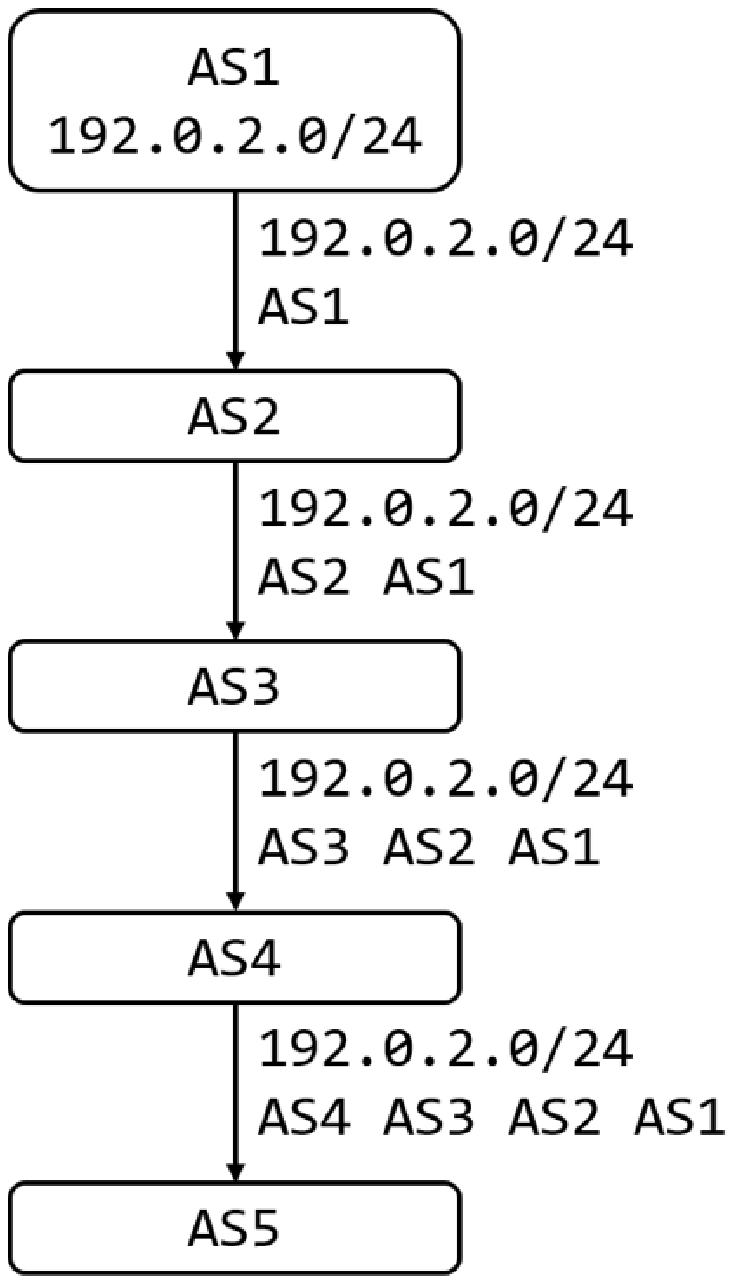}
 \end{center}
 \caption{Route Advertisement}
 \label{fig:RAdv}
\end{minipage}
\begin{minipage}{0.49\hsize}
 \begin{center}
\includegraphics[width=.99\textwidth, height=.20\textheight]{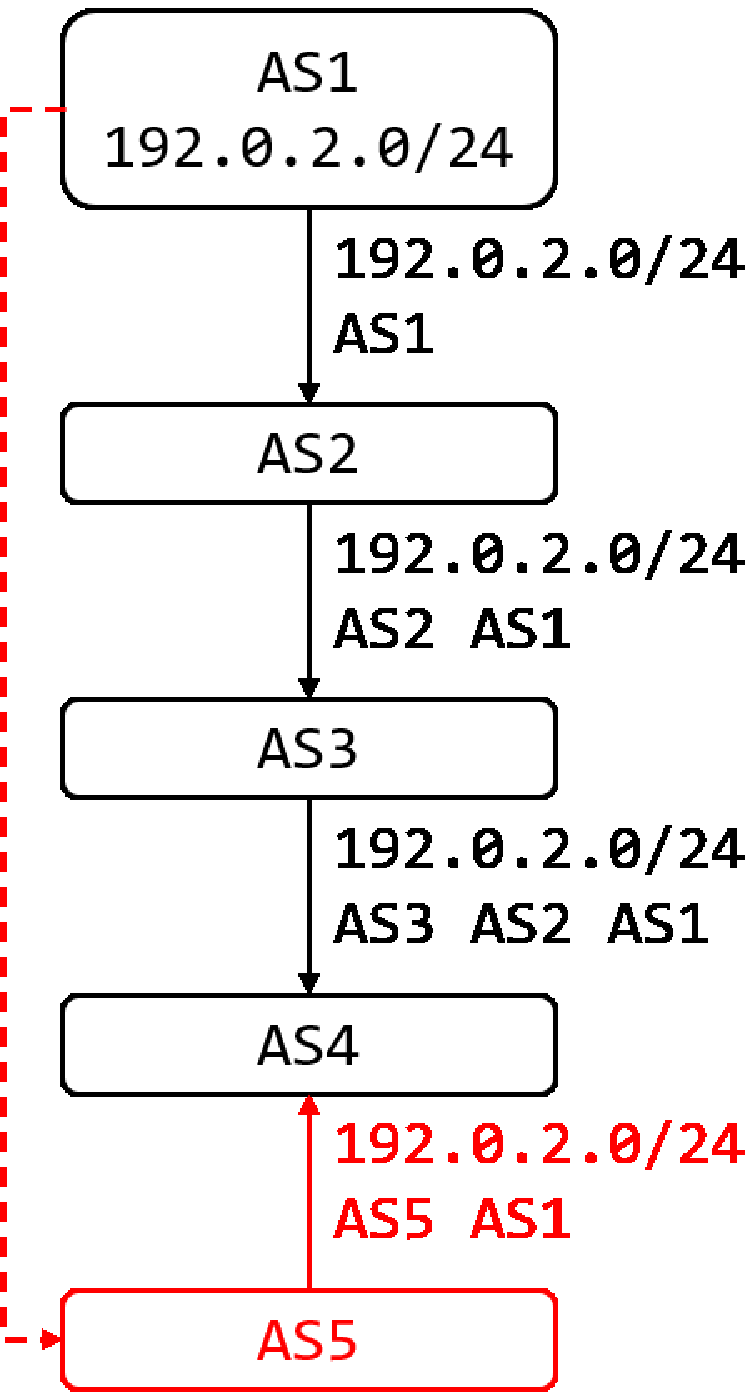}
 \end{center}
\caption{Example of Route Hijacking}
\label{fig:RHijack}
\end{minipage}
\end{figure}

Fig.~\ref{fig:RAdv} shows an example of a route advertisement on BGP. 
The route to AS1 with 192.0.2.0/24 registered in A4 is \texttt{AS3 AS2 AS1}. However, the \texttt{AS\_PATH} included in the update message can be intentionally rewritten by an AS to launch a route hijacking. 
As shown in Fig.~\ref{fig:RHijack}, AS5 advertises AS4 a shorter route to AS1, i.e., \texttt{AS5 AS1}. Despite the nonexistence of the route information, AS4 registers this information in its routing table according to the best path selection algorithm.    

BGPsec~\cite{lepinski_bgpsec_2017} prevents route hijacking by validating the \texttt{AS\_PATH} with the use of digital signatures on the routing information. Hereafter, we focus on the path validation provided by BGPsec. 

\begin{figure}[tbp]
 \begin{minipage}{0.49\hsize}
 \begin{center}
 \includegraphics[width=.99\textwidth, height=.20\textheight]{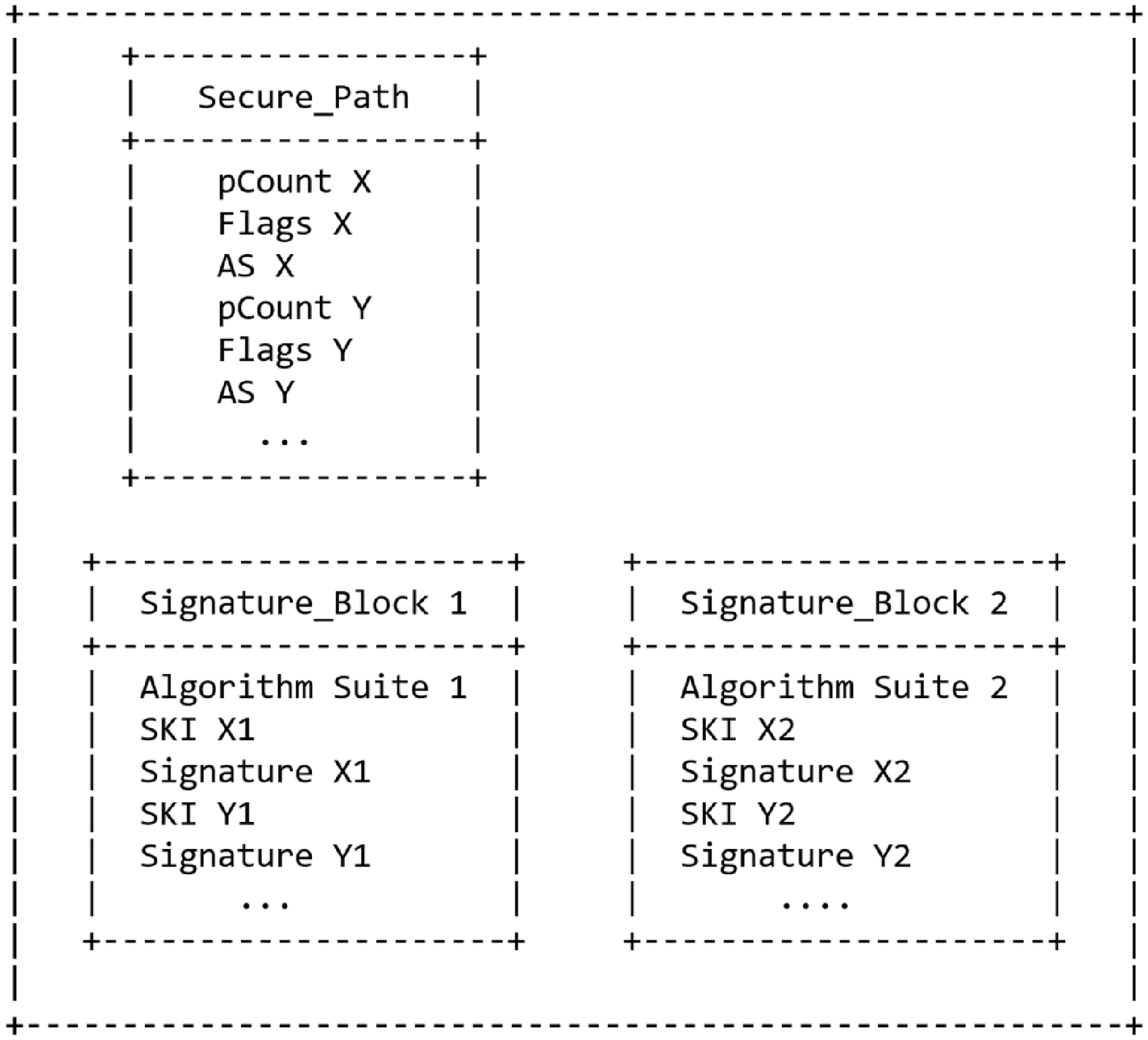}
 \end{center}
 \caption{Format of BGPsec\_PATH Attribute}
 \label{fig:SecPath}
\end{minipage}
\begin{minipage}{0.49\hsize}
 \begin{center}
\includegraphics[width=.99\textwidth, height=.20\textheight]{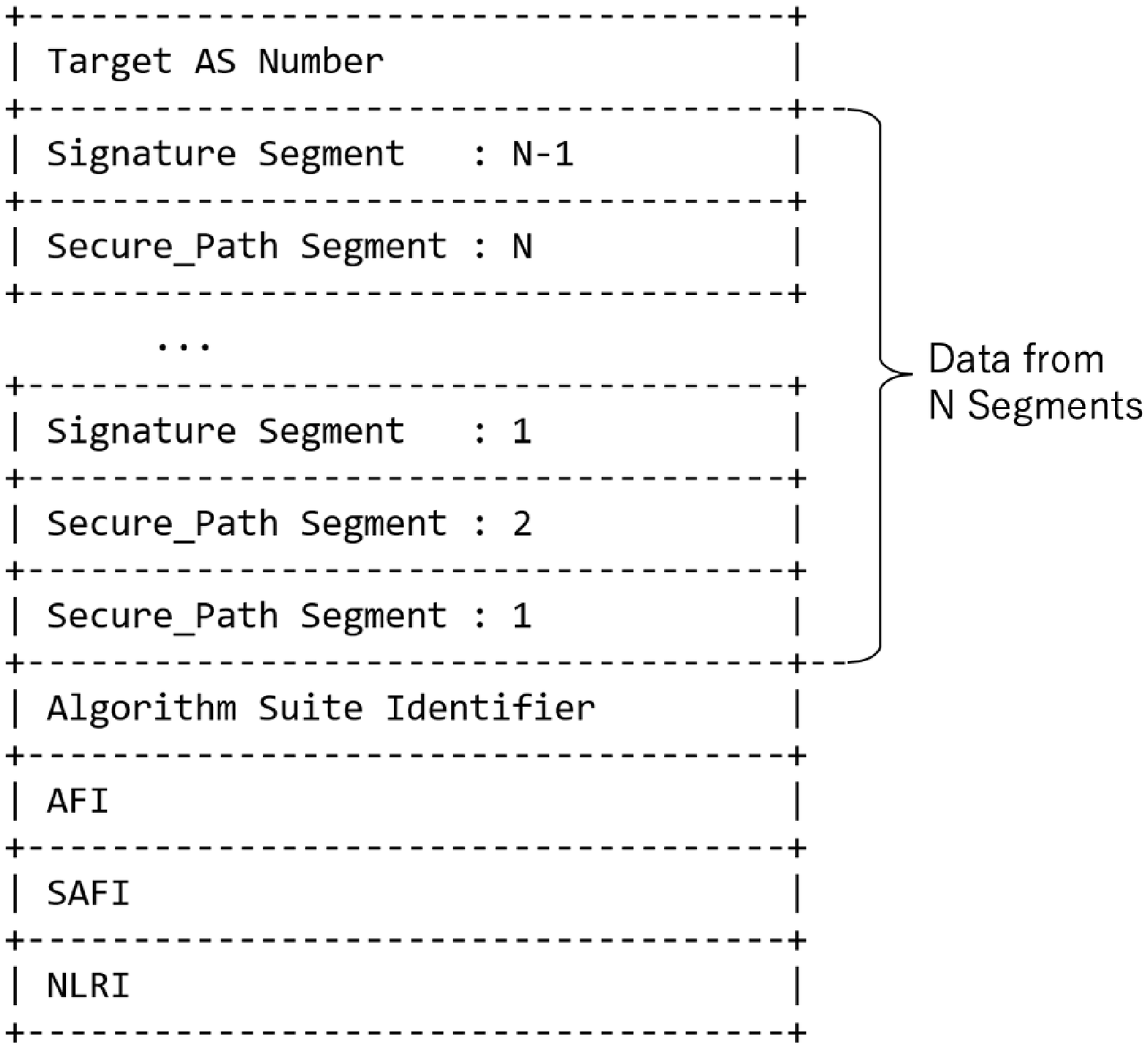}
 \end{center}
\caption{Sequence of Octets to Be Hashed}
\label{fig:HASH}
\end{minipage}
\end{figure}

\subsection{Path Validation}

In BGPsec~\cite{lepinski_bgpsec_2017}, \texttt{BGPsec\_PATH} attributes are defined instead of the \texttt{AS\_PATH} attributes of BGP. 
\texttt{BGPsec\_PATH} attributes contain \texttt{Secure\_PATH} and \texttt{Signature\_Block} as shown in Fig.~\ref{fig:SecPath}. 
\texttt{Secure\_PATH} lists the AS numbers of each AS that the routing information passed through, and it is identical to \texttt{AS\_PATH} of the conventional BGP. 
In contrast, \texttt{Signature\_Block} stores digital signatures generated by each AS specified in \texttt{Secure\_PATH}. 
The signature length varies according to the signature algorithm specified by \texttt{Algorithm Suite Identifier}.

Parameters of data strings to be signed are shown in Fig.~\ref{fig:HASH}. Each AS sends and receives an update message containing these parameters. We describe the important parameters below: 
\noindent \textbf{Target AS Number}: AS number of the destination of routing information. 


\noindent \textbf{Signature Segment}: Digital signatures, where the number of the signatures is identical to the number of ASes that the routing information passed through excluding the route origin.

\noindent \textbf{Secure\_Path Segments}: AS numbers of ASes that the routing information passed through, where at least the AS number of the route origin is required. 

\noindent \textbf{Algorithm Suite Identifier}: An identifier for specifying the signature algorithm used for signature generation. 



\noindent \textbf{NLRI}: Values of network addresses and their subnet mask managed by the route origin.

\subsection{Problem Setting} 

Since an update message on BGPsec contains digital signatures, the size of the update message balloons and a big part of which comes from the digital signatures. 
\begin{table}[tb] 
\centering
\caption{Existing Evaluation of Memory Size for BGPsec} 
\label{tab:Memoryize}
\hbox to\hsize{\hfil
\begin{tabular}{c c c c c}
\hline 
\multirow{2}{*}{} & \multicolumn{2}{c }{\texttt{BGP}} & \multicolumn{2}{c}{\texttt{BGPsec}} \\ \cline{2-5}
Year  & \shortstack{Number of\\ Paths}  & \shortstack{Memory\\ Size [GB]} & \shortstack{Number of\\ Paths}  & \shortstack{Memory\\ Size [GB]} \\ \hline
2020 & 6332177 & 0.13          & 6332177  &2.79\\ \hline 
2021 & 4433446 & 0.09          & 10130562 &4.47\\ \hline 
2022 & 2547235 & 0.05          & 14201374 &6.62\\ \hline 
2023 & 1149812 & 0.02          & 18111088 &7.99\\ \hline 
2024 & 355617   & 0.01          & 21794419 &9.61\\ \hline 
2025 & 0           & 0              & 25472541 &11.23\\ \hline 
\end{tabular}\hfil}
\end{table}
National Institute of Standards and Technology (NIST)~\cite{sriram_rib_2011} has shown the estimation results for memory size of routers on BGPsec. 
According to NIST, a BGP update message has an average size of 78 bytes, while a BGPsec update message is 388 bytes to 1188 bytes in size depending on the signature algorithms. 
We recall the result with ECDSA-256 in Table~\ref{tab:Memoryize}. 
In this table, the columns of Memory Size show total values with respect to routing tables registering destinations and those route attributes. 
Route information on BGPsec with respect to the world-wide level, i.e., full routes, requires a router to own more than 10 gigabytes of memory. 
Discussions and deployment of BGPsec have begun in 2016, and the deployment is estimated to finish in 2025. 
However, \textit{the complete deployment is nowhere in sight due to the memory size problem.}

\section{Bimodal Aggregate Signatures} \label{Bimodal Aggregate Signatures}

In this section, we discuss the bimodal aggregate signature scheme used as a new building block in APVAS. 
In our proposed scheme, when $n$ users generate $n$ individual signatures, these signatures can be aggregated in two ways, i.e., an interactive style of general aggregate signatures~\cite{boneh_aggregate_2003} and a signature-chain style of sequential aggregate signatures~\cite{LMRS04}. 

Compared with conventional aggregate signatures~\cite{boneh_aggregate_2003,LMRS04}, bimodal aggregate signatures can provide the security of BGPsec by virtue of signature chains as well as the efficiency of aggregating individual signatures even on individual paths. Specifically, the security of BGPsec depends on signature chains and only the sequential aggregate signatures provide such chains via signature aggregation. 
On the other hand, to improve efficiency, i.e., reduce memory consumption, signatures even on individual paths should be aggregated, and such capability is inspired only by signature aggregation of the general aggregate signatures. In other words, when either one of the conventional aggregate signatures is deployed, only one of either the security or the efficiency is guaranteed. 
In contrast, the use of bimodal aggregate signatures can solve the memory size problem without sacrificing the security. 
We show the algorithms of the bimodal aggregate signature scheme below. 

The proposed scheme is based on pairings defined as follows. 
Let $\mathbb{G}$ and $\mathbb{G}_T$ be groups with a prime order $p$. 
Then, a bilinear map $\textrm{\boldmath $e$}:\mathbb{G} \times \mathbb{G} \rightarrow \mathbb{G}_T$ is a map with the following conditions: 
for any $U, V \in \mathbb{G}$ and $a,b \in \mathbb{Z}_p^*$, 
$\textrm{\boldmath $e$}(aU, bV)= \textrm{\boldmath $e$}(U, V)^{ab}$ holds; 
for any generator $P \in \mathbb{G}$, 
$\textrm{\boldmath $e$}(P, P)\neq \textrm{\boldmath 1}_{\mathbb{G}_T}$ holds, where $1_{\mathbb{G}_T}$ is an identity element in $\mathbb{G}_T$;
and, for any $U, V \in \mathbb{G}$, $\textrm{\boldmath $e$}(U, V)$ can be computed efficiently. 
We assume that solving the discrete logarithm problems in $\mathbb{G}$ and $\mathbb{G}_T$ is computationally hard. 
We call the parameter $(p, \mathbb{G},\mathbb{G}_T,\textrm{\boldmath $e$})$ achieving the conditions above as \textit{pairing parameter}. 

\begin{algorithm}[b]
  \caption{Setup}
  \label{ams:setup}
  \begin{algorithmic}[1]
    \ENSURE 
	Public parameter $para$
    \STATE Generate pairing parameter $(p,\mathbb{G},\mathbb{G}_T,\textbf{e})$
    \STATE $P \leftarrow \mathbb{G} $
	\STATE Choose a hash function $H: \{0,1\}^* \rightarrow \mathbb{G}$ 
	\STATE $para=(p, \mathbb{G},\mathbb{G}_T,{\bf e}, P, H)$
  \end{algorithmic}
\end{algorithm}
\begin{algorithm}[b]
  \caption{UserKeyGen}
  \label{ams:ukeygen}
  \begin{algorithmic}[1]
	\REQUIRE Public parameter $para$ 
   \ENSURE Secret key $sk$, public key $pk$
    \STATE $x \leftarrow \mathbb{Z}_p$
	\STATE $X =xP$
	\STATE $sk= x$, $pk=X$
  \end{algorithmic}
\end{algorithm}
\begin{algorithm}[b]
  \caption{SeqAggSign}
  \label{ams:sign}
  \begin{algorithmic}[1]
    \REQUIRE public parameter $para$,  secret key $sk_{i}$, public key $pk_i$, plaintext $m_i \in \{0,1\}^*$, list $L=\{ (pk_j, m_j) \}_{j \in S }$ of public keys and plaintexts, signature $\sigma$
    \ENSURE Signature $\sigma$, list $L'=\{ (pk_j, m_j) \}_{j \in S } \cup \{ (pk_i, m_i) \}$ of public keys and plaintexts
	\IF{$L=\emptyset$} 
		\STATE set $\sigma = 0 $
	\ENDIF
    \STATE $c = H \left( \textbf{e} (\sigma, P) \parallel pk_i \parallel m_i \parallel_{j \in S_i} ( pk_j \parallel m_j ) \right) $
    \STATE $\sigma = \sigma + x \cdot H(c)$
  \end{algorithmic}
\end{algorithm}
\begin{algorithm}[b]
  \caption{AggSign}
  \label{ams:agg}
  \begin{algorithmic}[1]
    \REQUIRE public parameter $para$, list $L_1=\{ (pk_j, m_j) \}_{j \in S }$ of public keys and plaintexts, list $L_2=\{ (pk_j, m_j) \}_{j \in S' }$ of public keys and plaintexts, signature $\sigma_1$, signature $\sigma_2$
	\ENSURE signature $\sigma$, list $L' = L_1 \cup L_2$
    \STATE $\sigma = \sigma_1 + \sigma_2$
  \end{algorithmic}
\end{algorithm}
\begin{algorithm}[b]
  \caption{Verify}
  \label{ams:ver}
  \begin{algorithmic}[1]
    \REQUIRE public parameter $para$, list $L=\{ (pk_j, m_j) \}_{j \in S }$ of public keys and plaintexts, signature $\sigma$
    \ENSURE True or False
    \STATE For any $i \in S$, parse $pk_i $ as $X_i$
    \IF{all $(pk_i, m_i) \in S $ are distinct}
		\STATE {$\forall i, c_i = H \left( \left( \prod_{j \in S} \textbf{e} (c_j, X_i) \right) 
			 \parallel_{j \in S_i} ( pk_j \parallel m_j ) \right) $}
    		\IF{$\textbf{e} (\sigma, P) = \prod_{i \in S} \textbf{e} ( H ( c_i) , X_i) $}
			\RETURN True
		\ENDIF
	\ENDIF 
	\RETURN False
  \end{algorithmic}
\end{algorithm}

Hereafter, each signer is represented by a unique index $i$ in the algorithms for convenience. 
We denote by $S$ a set of signers for any signature and by $S_i$ a set of signers who join a chain of signatures from $1$ to $i$ for any $i$. 
We also denote by $\parallel$ a concatenation of any string and by $\parallel_{j \in S_i}$ concatenations of strings for any signer $j \in S_i$. 
We show the construction in {\bf Algorithms}~\ref{ams:setup}--\ref{ams:ver}. 
The fourth line of {\bf Algorithm}~\ref{ams:sign} and the third line of {\bf Algorithm}~\ref{ams:ver} are identical to a signature chain, i.e., $\sigma$ indicates a signature. 
More precisely, $\sigma$ indicates a signature, and $\textbf{e} (\sigma, P)$ in {\bf Algorithm}~\ref{ams:sign} and $\textbf{e} (c_j, X_i) $ in {\bf Algorithm}~\ref{ams:ver} are identical to computations which are the main core for verification of signatures. 
Intuitively, by inputting signatures and their verifications to a hash function, the same signature chains are constructed. 
Even after $\sigma$s themselves are aggregated and lost, the corresponding $\textbf{e} (c_j, X_i) $'s can be computed. Therefore, both a signature chain of sequential aggregate signatures and an interactive aggregation of general aggregate signatures can be provided simultaneously. 
The security of the proposed scheme can be proven formally under the computational Diffie-Hellman assumption in the random oracle model. We omit the details due to page limitation.


\section{Design of APVAS} \label{APVAS}

In this section, we present \textit{AS path validation based on aggregate signatures (APVAS)} as a new path validation protocol for BGPsec. 
As described in Section~\ref{Introduction}, APVAS is based on bimodal aggregate signatures. 
We first describe how bimodal aggregate signatures are deployed for path validation as a protocol specification. Then, we describe the prototype implementation on BIRD Internet routing daemon (BIRD), which is a software for BGP. 

\subsection{Protocol Specification}

\begin{figure}[tbp]
 \begin{minipage}{0.59\hsize}
 \begin{center}
 \includegraphics[width=.99\textwidth, height=.20\textheight]{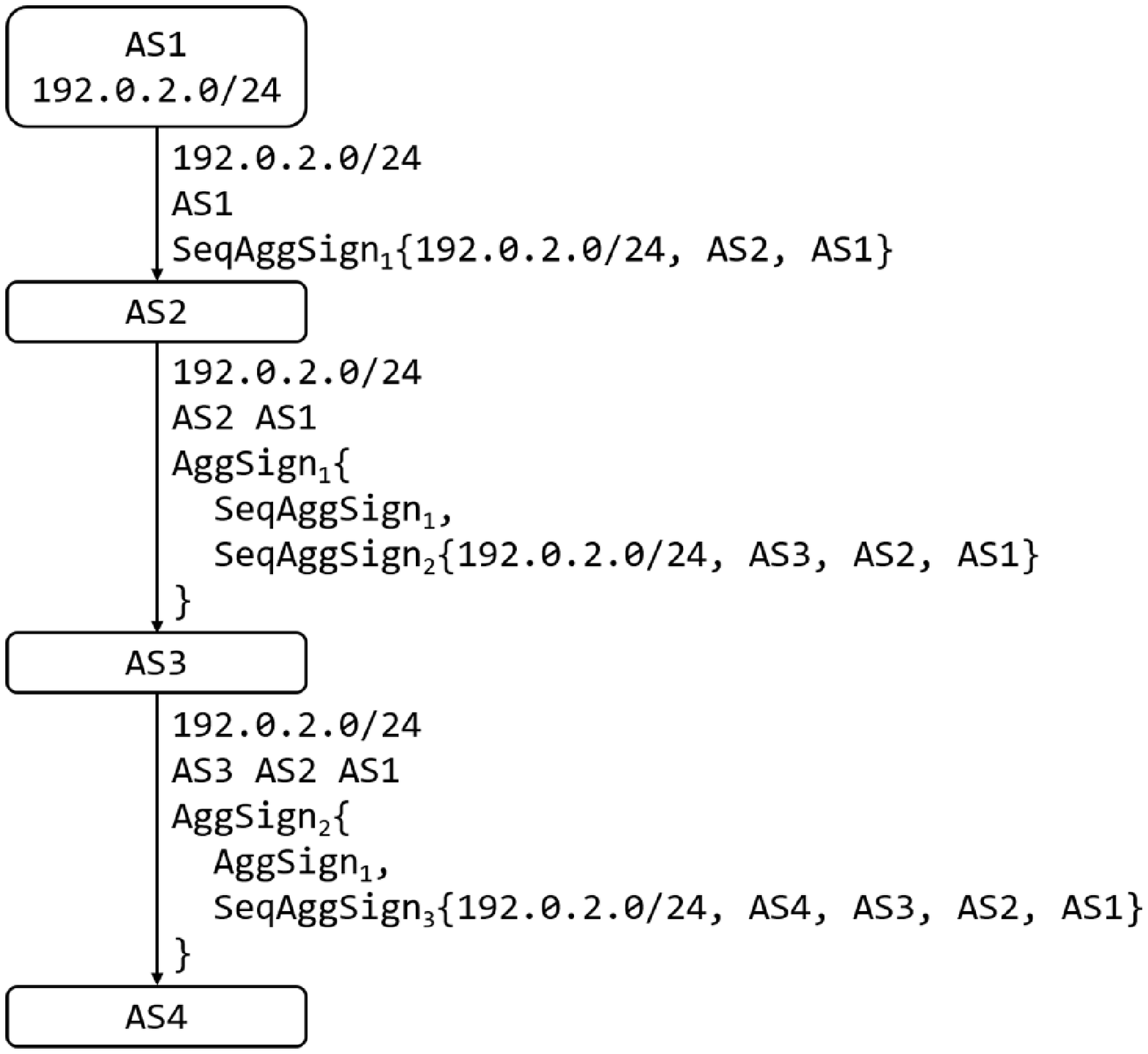}
 \end{center}
 \caption{Route Advertisement on APVAS}
 \label{fig:BGPsecAgg}
\end{minipage}
\begin{minipage}{0.4\hsize}
 \begin{center}
\includegraphics[width=.99\textwidth, height=.20\textheight]{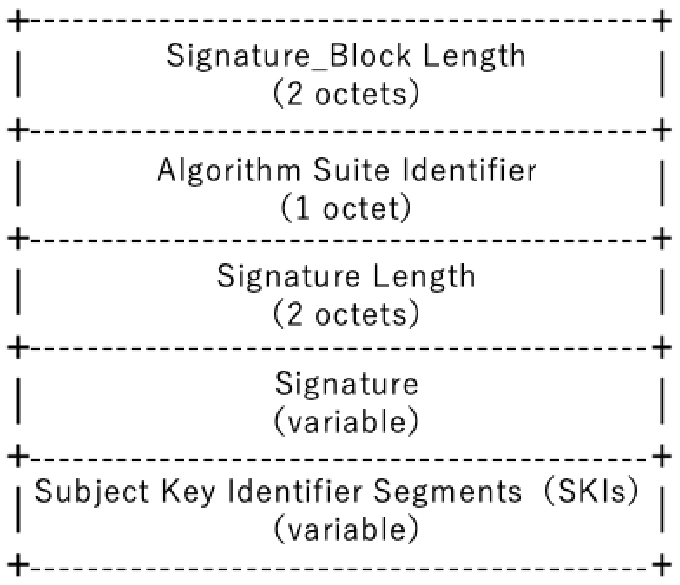}
 \end{center}
\caption{\texttt{Signature\_Block Format} on APVAS}
\label{fig:SigBlock}
\end{minipage}
\end{figure}

The specification of route advertisements on APVAS is shown in Fig.~\ref{fig:BGPsecAgg}. 
Each intermediate AS takes information from the received update message and then verifies the routing information with {\bf Algorithm}~\ref{ams:ver}. 
Then, for the received routing information and aggregate signatures, an AS generates a new aggregate signature with {\bf Algorithm}~\ref{ams:sign} and sends it to the neighbor ASes. 
Meanwhile, when an AS receives an update message including routing information from a different \texttt{ORIGIN}, the AS generates an aggregate signature as described above and then aggregates those signatures with {\bf Algorithm}~\ref{ams:agg}. 

\texttt{Signature Segment Format} defined in the BGPsec protocol contains \texttt{Subject Key Identifier (SKI)} with size of 20 bytes, \texttt{Signature Length} with size of 2 bytes, and \texttt{Signature} with a variable length. 
In contrast, to contain only a single signature in APVAS, \texttt{SKI} is defined as \texttt{SKI Segment} with size of 20 bytes. 
The new \texttt{Signature\_Block Format} of APVAS is shown in Fig.~\ref{fig:SigBlock}. 


The proposed scheme, i.e., {\bf Algorithms}~\ref{ams:setup}--\ref{ams:ver}, is utilized as the signature in APVAS as described above. 
We note that a plaintext $m$ corresponds to a received update message. 
More precisely, for data storing shown in Fig.~\ref{fig:HASH}, the remaining string except for \texttt{Signature Segment Format} is utilized as $m$. 
Likewise, for verification of update messages, the intermediate values of computation, i.e., outputs of bilinear maps, are utilized instead of a hashed value utilized for the previous signers to generate signatures because the update messages do not include the hashed values. 
Consequently, the size of update messages and the data size to be stored are reduced. 

\subsection{Prototype Implementation}

We implement a prototype of APVAS by extending the BGPsec-enabled BIRD~\cite{bird-bgpsec}, which is an implementation of BGPsec on BIRD, with a pairing library TEPLA\footnote{TEPLA: \url{http://www.cipher.risk.tsukuba.ac.jp/tepla/index.html}}. 

BIRD is a daemon software that utilizes a computer as a BGP router, and new functions can be introduced by writing in the C language. 
By editing a configuration file, we can set a router configuration and a network topology. 
The BGPsec-enabled Bird Routing Daemon~\cite{bird-bgpsec} is an extension of BIRD released by Secure Routing\footnote{Secure Routing: \url{http://www.securerouting.net/}}, 
and we implemented the prototype of APVAS by extending the BGPsec-enabled Bird Routing Daemon. 

Source codes related to BGP are in directories under \texttt{proto/bgp/}, and the codes to be improved are as follows. 
First, we improve encoding and decoding of update messages by modifying the \texttt{encode\_bgpsec\_attr()} function and the  \texttt{decode\_bgpsec\_attr()} function in \texttt{attrs.c}, respectively. 
We also improve signature generation and verification by replacing the \texttt{bgpsec\_sign\_data\_with\_key()} function and  the \texttt{bgpsec\_verify\_signature\_with\_key()} function in \texttt{validate.c}, respectively, with an implementation of the bimodal aggregate signatures with TEPLA. 
The parameters utilized in pairing computation are shown in Table~\ref{tab:parameter}. 
%
\begin{table}[tb] 
\caption{Version of Library and Utilized Parameters} 
\label{tab:parameter}
\hbox to\hsize{\hfil
\begin{tabular}{cc}
\hline \hline
TEPLA ver. & 2.0 \\ \hline
Pairing & ECBN254a \\ \hline
Finite Field &bn254\_fpa \\ \hline
P (Pre-generated) & 1462ea218754f628c4...\\ \hline
\end{tabular}\hfil}
\end{table}

\section{Experiments} \label{Experiments}

We created a prototype implementation of APVAS and conducted experiments on a virtual network with a simple topology to evaluate the memory consumption of APVAS and compare it with the results of the conventional method. 
The main purpose of our experiments is to take a first step in understanding the performance, i.e., in terms of memory consumption, of APVAS on actual devices.

\subsection{Experimental Environment}

While BIRD can deal with many kinds of network topologies, a linear network is the simplest network topology for understanding the relationship between the number of paths and the average \texttt{AS\_PATH} length. In such a topology, AS routers are connected with each other in a linear manner. Moreover, the number of paths advertised to the network is the total number of paths statically advertised by each AS, and the average \texttt{AS\_PATH} length is an average of distances between ASes which advertise the paths. 
In the experiment described below, we focus on a linear network to evaluate APVAS. 


Six virtual routers are configured under the machine environment shown in Table~\ref{tab:Environment}, and private AS numbers from 65001 to 65006 are assigned to these routers, respectively. 
Each router is assigned a static IP from 192.168.0.201 to 206 in accordance with the lowest digit of its own AS number, and the routers connect with each other to configure the network. 
For example, when AS65001 advertises routing information, the other ASes receive the same routing information. 
In doing so, the average \texttt{AS\_PATH} length is identical to a difference between AS numbers, making the evaluation of data easy.  
Since the upper bound for route advertisement whereby BIRD works stably is 250 according to our pre-experiments, we adopt the number of paths advertised by AS65001 from 50, 100, 150, 200, and 250, respectively. 
For instance, in a case where AS65001 advertises 200 paths to AS65004, the latter AS receives the 200 paths whose \texttt{AS\_PATH} length is 3. 
In the setting described above, the memory consumption for routers owning 200 paths with length 3 can be estimated. 


We evaluate the memory consumption of AS65002 to AS65006 based on the environment described above. 
Note that the process of obtaining a pair of secret and pubic keys is outside the scope of the experiment. 
We suppose that the pair of keys is generated and installed on each router in advance. Likewise, routing information for advertisement is randomly generated in advance. 


\begin{table}[tb] 
\caption{Environment} 
\label{tab:Environment}
\hbox to\hsize{\hfil
\begin{tabular}{cc}
\hline \hline
OS & Ubuntu 16.04 LTS \\ \hline
CPU & Intel Core i7-6500U \\ \hline
Memory & 1 GB \\ \hline
VMware ver. & Workstation Pro 14.1.2 \\ \hline
BIRD ver. & 1.6.0 \\ \hline
BIRD BGPsec ver. & 0.9 \\ \hline
\end{tabular}\hfil}
\end{table}

\subsection{Results}

\begin{figure}[t!]
\centering
\includegraphics[width=.50\textwidth, height=.20\textheight]{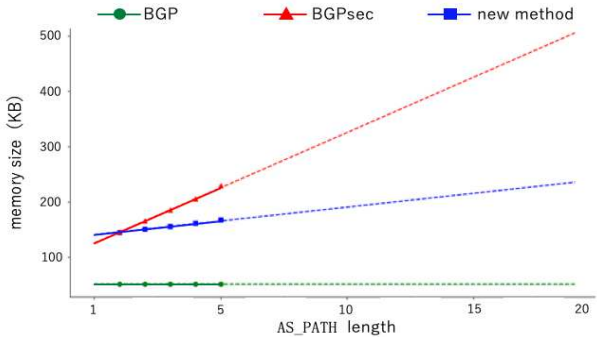}
\caption{Memory Size: The continuous lines are actual measurements on the experiments while the dotted lines are estimations from the measurements.}
\label{fig:MemGraph}
\end{figure}

The memory consumption when each AS receives 200 paths is shown in Table~\ref{tab:MemSize}. The table shows only the memory size related to routing tables and route attributes, although BIRD includes the routing tables, route attributes, Route Origin Authorization (ROA) table, and the protocol information. 
APVAS is successful in reducing the memory size in comparison with the conventional BGPsec because the sizes of other information in the current specification~\cite{lepinski_bgpsec_2017}, e.g., ROA tables and the protocol itself, are stable in our experiment. 
The results of the memory consumption under the same setting for BGP routers and the conventional BGPsec routers are shown in Table~\ref{tab:MemSize_BGP} and Table~\ref{tab:MemSize_BGPsec}, respectively, and the entire data is visualized in the continuous lines of Fig.~\ref{fig:MemGraph}. 


The results show that the memory size of APVAS is smaller than that of the conventional BGPsec. 
The memory size of APVAS becomes larger at the average \texttt{AS\_PATH} length 1 because of the data size of the bimodal aggregate signatures. 
More precisely, while the conventional BGPsec utilizes ECDSA with bit length of 384 bits per signature, the bimodal aggregate signatures have bit length of 512 bits. 


\begin{table}[tb] 
\caption{Memory Size of APVAS (with 200 Paths)} 
\label{tab:MemSize}
\hbox to\hsize{\hfil
\begin{tabular}{cccccc}
\multicolumn{6}{r}{Memory Size [KB]}\\
\hline \hline
AS Number & 65002 & 65003 & 65004 & 65005 & 65006 \\ \hline
Routing Table & 46 & 46 & 46 & 46 & 46 \\ \hline
Route Attribute & 116 & 122 & 128 & 134 & 140 \\ \hline
\end{tabular}\hfil}
\end{table}

\begin{table}[t] 
\caption{Memory Size of BGP Routers (with 200 Paths)} 
\label{tab:MemSize_BGP}
\hbox to\hsize{\hfil
\begin{tabular}{cccccc}
\multicolumn{6}{r}{Memory Size [KB]}\\
\hline \hline
AS Number & 65002 & 65003 & 65004 & 65005 & 65006 \\ \hline
Routing Table & 46 & 46 & 46 & 46 & 46 \\ \hline
Route Attribute & 10 & 10 & 10 & 10 & 10 \\ \hline
\end{tabular}\hfil}
\end{table}

\begin{table}[t] 
\caption{Memory Size of BGPsec Routers (with 200 Paths)} 
\label{tab:MemSize_BGPsec}
\hbox to\hsize{\hfil
\begin{tabular}{cccccc}
\multicolumn{6}{r}{Memory Size [KB]}\\
\hline \hline
AS Number & 65002 & 65003 & 65004 & 65005 & 65006 \\ \hline
Routing Table & 46 & 46 & 46 & 46 & 46 \\ \hline
Route Attribute & 104 & 124 & 144 & 164 & 188 \\ \hline
\end{tabular}\hfil}
\end{table}

\subsection{Consideration}

According to the empirical study of Wang et al.~\cite{wang2016inferring}, the average \texttt{AS\_PATH} length on the Internet is about 3.9 and the longest path is  about 20. When the average \texttt{AS\_PATH} length is 4, the size of route attributes by APVAS became 80\% less than the conventional BGPsec. The bulk of memory consumption is related to the size of route attributes~\cite{sriram_rib_2011}, and thus the results above confirm that APVAS can reduce memory consumption by 80\%. Furthermore, the memory consumption until length 20  can  be  estimated by the least squares method based  on  the experimental results  as  shown  on  the  dotted lines  in Fig.~\ref{fig:MemGraph}. In proportion to the length of \texttt{AS\_PATH}, the difference in the memory consumption between APVAS and the conventional BGPsec will become larger because the memory size of APVAS is stable from the aggregate signatures via {\bf Algorithm}~\ref{ams:agg} even if different route attributes appear. As shown in Table~\ref{tab:Memoryize}, the memory consumption of BGPsec is expected to increase in the future, and APVAS can be a practical solution. We are in the process of creating a detailed discussion on the performance of APVAS. 

\section{Conclusion} \label{Conclusion}

In this paper, we proposed APVAS, a path validation method that deploys novel bimodal aggregate signatures to reduce memory consumption on BGPsec. 
We implemented a prototype of APVAS via BIRD and measured the memory consumption with actual routers. Our experimental results confirm that APVAS can reduce memory consumption by 80\% in comparison with conventional BGPsec. 
We are preparing for the release of our implementation of APVAS to encourage development of BGPsec and future works.
We hope that APVAS will serve as a base for future studies on BGPsec.

We plan to experiment further with full routes as future work. 
Although the experiments in this paper were conducted on a linear network as the simplest topology as the first step of a long way to practical realization, the performance in an environment with full routes should be clarified before deployment in the world-wide level. 
Furthermore, we discovered a new problem where the throughput of routers is downgraded by the computational cost of bimodal aggregate signatures, which is heavier than that of ECDSA. 
Thus, further studies on reducing not only the memory consumption but also the computational cost, which takes misconfiguration such that a large amount of routing information is advertised into account, will need to be undertaken. 



\section*{Acknowledgement}
This research was supported in part by the Japan Society for the Promotion of Science KAKENHI Numbers 18K18049 and the Secom Science and Technology Foundation.

\bibliographystyle{unsrt}
\bibliography{reference}

\end{document}